\def\comment#1{}

\newcommand{\ve}[1]{\mathbf{#1}}

\newcommand{\im}{{\rm i}}
\newcommand{\beg}{\begin{eqnarray}}
\newcommand{\eee}{\end{eqnarray}}

\documentclass[prl,letterpaper,twocolumn,aps,epsf]{revtex4}
\usepackage{dcolumn}
\usepackage{amsmath}
\usepackage{graphicx}%
\def\cm#1{}

\begin{document}
%\title{Nuclear hydrogenic superconductivity in multi-component quantum fluids}
\title{Observability of a projected new state of matter: a metallic superfluid}

\author{Egor Babaev${}^{1,2}$, Asle Sudb{\o}${}^2$, N. W. Ashcroft${}^1$}

\affiliation{
${}^1$Laboratory of Atomic and Solid State Physics, Cornell University,  Ithaca, NY 14853-2501 USA
\\ ${}^2$ Department of Physics, Norwegian University of Science and Technology, N-7491 Trondheim, Norway 
}

\begin{abstract}
Dissipationless quantum states, such as superconductivity and superfluidity, have 
attracted interest for almost a century. A variety of systems  exhibit 
these macroscopic quantum phenomena, ranging from superconducting electrons in 
metals to superfluid liquids, atomic vapours,  and even large  nuclei. It was recently 
suggested that liquid metallic hydrogen could form two new unusual  dissipationless 
quantum states, namely the {\it metallic superfluid} and the {\it superconducting 
superfluid}. Liquid metallic hydrogen is projected to occur only at an extremely high 
pressure of about $400$GPa, while pressures on hydrogen of 320 GPa having already 
been reported. The issue to be adressed is if this state could be experimentally 
observable {\it in principle}. We propose four experimental probes for detecting it.
\end{abstract}

\pacs{71.10.Hf, 74.10.+v, 74.90.+n}

\maketitle
\newcommand{\la}{\label}
\newcommand{\aaa}{\frac{2 e}{\hbar c}}
\newcommand{\Pfaff}{{\rm\, Pfaff}}
\newcommand{\kA}{{\tilde A}}
\newcommand{\G}{{\cal G}}
\newcommand{\cP}{{\cal P}}
\newcommand{\M}{{\cal M}}
\newcommand{\E}{{\cal E}}
\newcommand{\btd}{{\bigtriangledown}}
\newcommand{\W}{{\cal W}}
\newcommand{\X}{{\cal X}}
\renewcommand{\O}{{\cal O}}
\renewcommand{\d}{{\rm\, d}}
\newcommand{\bfi}{{\bf i}}
\newcommand{\e}{{\rm\, e}}
\newcommand{\bfx}{{\bf \vec x}}
\newcommand{\bfn}{{ \vec{\bf  n}}}
\newcommand{\bfs}{{\vec{\bf s}}}
\newcommand{\bfE}{{\bf \vec E}}
\newcommand{\bfB}{{\bf \vec B}}
\newcommand{\bfv}{{\bf \vec v}}
\newcommand{\bfU}{{\bf \vec U}}
\newcommand{\bfp}{{\bf \vec p}}
\newcommand{\f}{\frac}
\newcommand{\bfA}{{\bf \vec A}}
\newcommand{\non}{\nonumber}
\newcommand{\be}{\begin{equation}}
\newcommand{\ee}{\end{equation}}
\newcommand{\ba}{\begin{eqnarray}}
\newcommand{\ea}{\end{eqnarry}}
\newcommand{\bastar}{\begin{eqnarray*}}
\newcommand{\eastar}{\end{eqnarray*}}
\newcommand{\half}{{1 \over 2}}
Historically, experimental discoveries of new quantum
fluids have often had impact well beyond the physics of condensed matter. The most important
quantum fluid states are: 
{\bf 1}) superconductivity in metals (1911), {\bf 2}) superfluidity in ${}^4He$ (1937),
 {\bf 3}) superfluidity in ${}^3He$ (1972),  {\bf 4})
high-$T_c$ $d$-wave superconductivity in the copper oxides (1986), and {\bf 5})
Bose-Einstein condensation of ultracold atoms confined in optical traps (1995).
We may also mention recent experiments centered on finding a supersolid state 
in ${}^4He$ \cite{chan}, which, if confirmed, would add crystalline solids to 
the list of substances with ``super" properties along with liquids, vapors and 
electrons in metals. 

Most of these experimental discoveries required novel theoretical ideas  for their 
interpretations, which  eventually inspired a number of corresponding novel notions 
in other branches of physics. A notable example is the seminal work of Bardeen, Cooper, 
and Schrieffer providing a theory of conventional phonon-mediated superconductivity, 
which influenced the appearance of the Nambu-Jona-Lasinio model describing dynamical 
symmetry breaking in particle physics  \cite{NJL}. The phase and spin degrees of freedom 
in neutral superfluids are naturally related to Goldstone bosons. The Meissner effect in 
superconductors is a counterpart to the Higgs effect, while the Abrikosov vortices in 
superconductors form counterparts to Nielsen-Olesen cosmic strings \cite{NO}.  There are 
numerous other examples of deep and intriguing connections between physical phenomena 
taking place on the macro- and micro-scales \cite{seealsovolovik}. This illustrates rather 
strikingly how Nature appears to operate on similar principles on vastly different energy- 
and length scales, and especially how experimental advances in condensed matter physics 
can indirectly influence and inspire ideas relevant to other branches of physics.

A reasonable question to raise is where further experimental advances in the field of 
quantum fluids might yet arise. An intriguing possibility,  which now appears to be 
experimentally realizable due to a breakthrough in the synthesis of  ultrahard diamonds 
\cite{Hemley}, is the low-temperature liquid metallic state of hydrogen (LMH). As shown 
originally by Heitler and London, the substantial homonuclear bond in molecular hydrogen 
owes its existence to the symmetric form of the two-electron wave-function (the spin 
function being antisymmetric). However, in a condensed state and under the action of 
compression, the electronic charge density associated with this bond (and corresponding 
pair potential) is expected to be systematically transferred from the intra-molecular 
regions, to the inter-molecular regions. A weakening both of the intra-molecular 
potential and the short-range (repulsive) part of the inter-molecular potential is then
anticipated. Since these are the interactions which ultimately lead to spatial order, and 
since there is a concomitant rise in zero-point energies with compressions, it is also 
to be expected that the melting point will decline, indeed an effect recently reported by 
Bonev {\it et. al.} \cite{Bonev} in very extensive simulations. Further, the continued 
transference of electron density into the interstitial regions also carries with it the 
possibility that (exactly as in ${}^4He$ under ordinary conditions) there may be a range 
of densities where the ordering energies become quite minor compared to zero-point energies. 
In this situation the result may be a ground-state liquid as the preferred phase.
Importantly, it will also be metallic, because densities are sufficient to induce 
an insulator-metal transition en route. Such a metallic liquid at temperatures of 
order 100K is expected to form Cooper pairs of electrons \cite{ja}. At lower 
temperatures also protonic Cooper pairs are expected to form \cite{NWA2}. In liquid 
metallic deuterium, the deuterons are spin-one bosons which should likewise lead 
to Bose condensate with no pairing mechanism required. 

 It was  recently demonstrated 
that once deuterons or Cooper pairs of protons are present along with Cooper pairs of 
electrons, the resulting ``super"-state does not then fall into any class of existing 
quantum fluids \cite{Nature}.

Two key aspects of this two-component condensate are: ({\it i}) It features a
{\it superfluid} mode of co-directed currents of protons and electrons which supports a 
superflow of mass but no charge transfer, and a {\it superconducting mode} of counter-directed
currents of protons and electrons involving dissipationless transfer of charge as well 
as mass \cite{frac}. 
({\it ii})
The neutral superfluid mode does not couple to an external magnetic field, while the charged 
superconducting mode does. Moreover, the neutral and charged modes are subject to topological 
constraints originating with condensates that are described by complex scalar 
fields whose phases should be singlevalued. Thus, topological defects in the 
form of vortices have a topological charge both in the superconducting and superfluid sectors of 
the model  \cite{frac}. In this system,  superconducting and superfluid properties are therefore 
inextricably intertwined. This has  numerous physical consequences \cite{Nature},\cite{fm}-\cite{KT}, 
one such being that if the system features superconductivity of type-II, then 
by applying an external magnetic field and controlling temperature, one should be able to drive 
the system through various topological phase transitions where it can aquire {\it selectively} 
either superconducting or superfluid properties \cite{Nature}. The mechanisms and universality 
classes of some of these phase transitions have recently been studied in considerable 
detail \cite{ssbs,prb}. 

A general $N$-component mixture of individually conserved condensates at low temperatures should 
be described by the following $N$-component  Ginzburg-Landau model, with a free-energy
%\begin{widetext}
\begin{equation}
F = \left \{ \sum_{\alpha=1}^N \frac{1}{2}{|(\nabla -\im
e\ve{A})\psi^{(\alpha)}|^2}
+ V(\{\psi^{(\alpha)}\})\right \} + F_{\ve A}.
\label{GL}
\end{equation}
Here, $F_{\ve A} = \int d \ve r \frac{1}{2}(\nabla\times\ve{A})^2 $ and the condensate 
masses $M^{(\alpha)}$  have been absorbed in the amplitudes 
$|\psi^{(\alpha)}|^2 = |\Psi^{(\alpha)}|^2/M^{(\alpha)}$ 
for notational simplicity. The condensate order parameters are complex fields denoted 
by $\Psi_\alpha = |\Psi_\alpha|e^{i \theta^{(\alpha)}}$, where $\alpha = 1..N$, and 
${V} (|\Psi_\alpha|^2) 
%$ is only a function of $|\Psi_\alpha|^2$ 
=b_\alpha |\Psi_\alpha|^2+\frac{c_\alpha}{2}\
|\Psi_\alpha|^4+d_{\alpha\beta}|\Psi_\alpha|^2|\Psi_{\beta}|^2$. %$
Note the absence of Josephson coupling between different condensate components. This is an important 
consequence of the fact that each individual condensate is conserved: Cooper-pairs of electrons 
cannot be converted into Cooper-pairs of protons and vice versa.  Moreover, the model Eq. (\ref{GL}), 
where all condensates are taken to be  $s$-wave, should be sufficient to capture the essential 
physics involved in the four proposaled experiments to be described below. We can also exclude pairing 
of different hydrogenic nuclei when we have very different Fermi momenta. The model in Eq. (\ref{GL}) 
is invariant under change of sign of a charge of any component $(e \to -e)$ with a simultaneous sign 
change in phase $(\theta^{(\alpha)} \to -\theta^{(\alpha)})$. Hence, we can choose the representation 
where all fields have the same charge sign, but the phase of  a positively charged
condensate is multiplied  by $-1$. The {\it superfluid properties} of the model are then revealed 
when the variables in Eq. (\ref{GL}) are separated into gradients of phase differences which 
do not couple to the vector potential and represent neutral modes, and  a sum of all phases 
coupled to vector potential which is a charged mode (for details see \cite{frac},\cite{Nature}, 
\cite{sss}-\cite{KT}). Accordingly (\ref{GL}) can be rewritten
%\end{widetext}
%\begin{widetext}
\begin{eqnarray}
F &=&  \biggl\{   
  \frac{1}{4 \Psi^2} \biggl[ \sum_{\alpha,\beta=1}^N |\psi^{(\alpha)}|^2  |\psi^{(\beta)}|^2 
 \biggl( \nabla (\theta^{(\alpha)}- \theta^{(\beta)}) \biggr)^2
 \biggr]+ \nonumber \\ && \biggl( 
\frac{1}{2 \Psi^2}  \sum_{\alpha = 1}^N |\psi^{(\alpha)}|^2 \nabla \theta^{(\alpha)}  - e \Psi^2 \ve A 
\biggr)^2  \biggr\} 
+ F_{\ve A},
\label{GLS}
\end{eqnarray}
%\end{widetext}
where $\Psi^2=\sum_{\alpha=1..N}|\psi^{(\alpha)}|^2$.
We shall base our discussion below largerly on
the existence of neutral and charged modes which are
explicitly identified in (\ref{GLS}).

Along with the experimental challenge of achieving the high pressures required to induce 
hydrogen to take up a liquid  metallic state, a central question is  whether it is possible 
to confirm experimentally the very existence of the liquid metallic state itself, i.e. what
would be experimeally accessible manifestations in the protonic superconductivity which 
is expected to coexist with electronic superconductivity at low temperatures. The main
difficulties associated with observing such a state are the following:
(i) the system is confined 
in a high-pressure diamond anvil cell of small dimension; (ii) protonic superconductivity
cannot be probed even in principle with conventional external electronic contacts simply 
because protons would not enter the contacts. This rules out resistivity measurements. 
(iii) Because the critical temperature for electrons is expected to be much higher 
than that for protons, another standard experimental technique, namely measurement of 
the Meissner effect, is also inapplicable for detecting protonic superconductivity.

Nonetheless, we will point out several possibilities of experimentally probing and 
confirming the presence of protonic superconductivity in a high pressure diamond 
cell containing LMH or a mixture of the hydrogen isotopes. The  protonic superconductivity 
and superfluidity detection experiments suggested below are all based on exploiting the 
topological properties of the $U(1)\times U(1)$ or general $[U(1)]^N$ condensate and 
therefore do not depend principally on microscopic details.

First, we comment on the manner in which a ground-state or near ground-state liquid phase 
of metallic hydrogen, or deuterium, may be unambiguously identified. Given the recent 
advances in neutron beam focusing (focused beams as small as 100$\mu$ are possible at 
present), direct structural probing (especially for deuterium) from samples confined 
in diamond cells may be a relatively obvious route. For both hydrogen and deuterium 
the spin of the neutron might also be usefully engaged in the probing of magnetic order 
on a span of length scales. Next, we propose four possible experimental probes of 
protonic superconductivity and superfluidity.

{\bf 1.}{\bf Quench induced temperature-dependent fractional magnetic flux.} Perhaps the most 
straightforward method of unequivocally confirming the presence of protonic superconductivity 
in a diamond anvil cell {\it which might even allow measurement of a protonic gap}, is to produce 
a multiply connected physical space as shown in Fig. 1, the LMH  then occupying a torus. Thermal 
quench (rapid cooling through the superconducting transition) in a multiply connected space 
will in  general result in  non-trivial phase windings of the condensates \cite{zurek}. This 
will result in a trapping of quench-induced magnetic flux, given by the expression \cite{frac}
\beg
\Phi=\frac{\frac{|\Psi_e(T)|^2}{m_e}n_e-\frac{|\Psi_p(T)|^2}{m_p}n_p}
{
\frac{|\Psi_e(T)|^2}{m_e}+\frac{|\Psi_p(T)|^2}{m_p}}\Phi_0
\label{fl}
\eee
where $n_{e,p}$ are the quench-induced windings of protonic and electronic condensates 
phases in units of $2\pi$, respectively and $\Phi_0$ is the magnetic flux quantum.
(In general some corrections arising from the Andreev-Bashkin effect
\cite{Andreev} might be expected to enter this expression, but they will not
alter the picture in any essential way.)
 Further, if 
for example, a rapid cooling through the $T_c^p$ produces the windings $n_e=0,n_p=1$ 
then at low temperature the  flux would be of order of $10^{-3} \Phi_0$ which is at 
least three orders of magnitude larger than the maximum flux resolution in modern 
experiments. An important point is that the critical temperature $T_c^e$ for electrons 
is expected to be much higher than the critical temperature for protons $T_c^p$.
Thus, at temperatures of the order of $T_c^p$, we have $|\Psi_e(T)| \approx |\Psi_e(0)|$. 
The flux will be controlled  by the temperature dependent density of the 
protonic condensate. This should allow a very accurate determination of the temperature 
dependence of the protonic gap and $T_c^p$ by measuring the confined fraction of magnetic 
flux quantum (\ref{fl}). We also stress that thermal quench could be produced by 
irradiation, illumination and by variation of pressure. For this probe it is not  
important that the condensates be of type-I or type-II. This technique is also 
applicable to mixtures of $N$ condensates.  Then, 
if only the condensate $\Psi_{\gamma}$ has a $2\pi$ phase winding the quench-induced 
flux will be 
$
 \Phi=%\oint {\bf A} d {\bf l} =
\frac{|\Psi_{\gamma}(T)|^2}{m_{\eta}}
\left[ \sum_{\alpha=1}^N\frac{
|\Psi_{\alpha}(T)|^2}{m_{\alpha}}
\right]^{-1} \Phi_0 $.
%We note also that there might arise corrections to this expression because of 
%the Andreev-Bashkin effect \cite{andr} which however are not of major importance.

% \begin{widetext}
\begin{center}
\begin{figure}[htb]
\centerline{\scalebox{0.3}{{\includegraphics{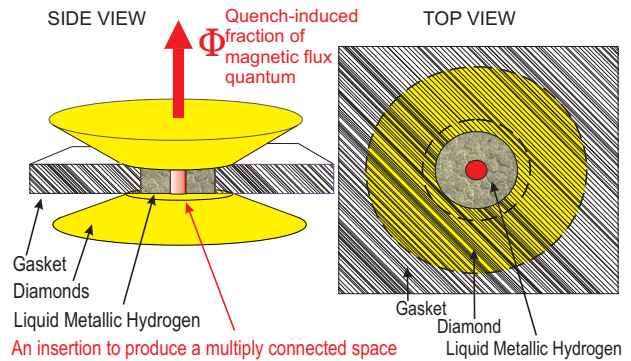}}}} 
\caption{(Color online) Liquid metallic hydrogen in a high pressure diamond anvil cell. The red insertion 
makes the sample multiply connected. Thus, a thermal quench should produce $2\pi\times$[integer] 
winding of the phase of the protonic condensate. This will result in a detectable fractional 
magnetic flux emitted from the diamond anvil cell. The fraction of the flux quantum will 
depend on the ratios of superfluid densities of electronic and protonic condensates.}
\end{figure}
\end{center}
%\end{widetext}
{\bf 2.} {\bf Fractionally quantized magnetic field induced by rotation}. 
The existence of a superfluid mode in the system means that a rotation
 should produce a vortex lattice 
in a similar way as in neutral systems, such as rotating buckets of superfluid helium ($^4He$). The 
fact that a vortex in such a system features both neutral vorticity and carries magnetic flux 
(\ref{fl}), means that  there will be potentially detectable  {\it rotation-induced 
vortices carrying a magnetic flux}.  We point out that since we are speaking of rotation-induced 
vortices in a neutral superfluid mode (although the vortices also carry a magnetic flux), for 
this probe it is also of  no importance that the system be a superconductor of type-I or type-II, 
or even of a mixed  type-I/type-II type \cite{nt2}. A difficulty in realizing such an experiment 
in presently available diamond anvil cells is their small dimensions and the low mass of electronic 
Cooper pairs which makes the critical rotation frequency very high. 
However, we mention a recent breakthrough in fabricating diamonds with 
large dimensions and fewer defects \cite{Hemley}.

{\bf 3.} {\bf Magnetization jump in a transition to a mixed type-I and type-II superconducting 
state}. If both the protonic and electronic condensates at 
low temperature are type-II superconductors, then the external field measurements might reveal 
a particular physical signature which is {\it qualitatively} different from the  behaviour 
of a single gap system, and which thus can also be used to confirm the presence of a superconducting 
state of protons. Upon heating the system close to the critical temperature of the protonic condensate, 
the coherence length for protons should start to diverge while the magnetic penetration length 
will not vary significantly since it is controlled by the electronic condensate, and given by 
$\lambda = 1/e (|\Psi_e|^2/m_e + |\Psi_p|^2/m_p)^{-1/2}$. Thus, there will  necessarily  exist a 
temperature range where the coherence length of the protonic condensate will be  larger than 
$\lambda$. However, in such a situation the vortices can nonetheless be thermodynamically stable 
\cite{nt2}. It follows that, if the electrons form a type-II condensate, such a situation may 
lead to a conversion of the phase transition in an external field  from second to first order with a 
jump of magnetization \cite{nt2}. The jump in magnetization will be controlled by the 
non-monotonicity in the interaction between vortices. The longer range attractive part of the 
intervortex interaction originates with the proton core with the characteristic energy per unit 
length involved being of order $E_p^c \times \xi_p^2$, where $E_p^c$ is the protonic condensation 
energy and $\xi_p$ is the protonic coherence length. However, this effect would  be eliminated if 
the system enters a sublattice vortex liquid state \cite{Nature,ssbs} at lower temperature, which 
we expect would be the more likely scenario. Another possibility of detection of a magnetization 
jump is the following: It is expected that the  parameters of the electronic superconductivity 
could be tuned by applying pressure. In particular, they might possibly be tuned in a wide range 
from type-I to type-II \cite{ja}. If the protonic coherence length at temperatures lower than $T_c^p$ 
is much smaller than that of the electronic condensate,
a pressure-induced crossover from type-II to type-I superconductivity 
in electrons could conceivably lead to a detection of a magnetization jump. From the magnetization 
jump and its temperature and pressure dependence one can then extract data on the order parameters. 

{\bf 4.} {\bf A two-dimensional flux noise probe.} If an experiment is conducted in a quasi-2d geometry
then this system may undergo a Kosterlitz-Thouless (KT) transition \cite{KT} which may be 
detectable in flux-noise measurements. Flux detection coils have already been effectively utilized 
in high pressure experiments. In particular small coils can be formed inside
                         artifically grown diamonds and  hence such a measurement 
                         may also be feasible on LMH, were it  to be realized. 
It is expected that a protonic superconductor will  
undergo a BEC-BCS crossover with increasing pressure. There will therefore exist a pressure 
range where the temperature of KT transition temperature would be significantly
lower than the temperature of  thermal Cooper pair decomposition. An advantage of this
particular  probe 
is that the two-component system undergoes a KT transition even in type-I limit \cite{KT}. 
The flux noise measurements in principle yield detailed information on vortex interactions, 
and therefore also on the possible 
existence of  a composite neutral mode or multiple
composite neutral modes in the case of a mixture
of hydrogen isotopes \cite{prb}.
The disadvantage of this method is the necessity of a quasi-2d geometry
and the finite sample size limitations that this imposes.

In summary, liquid metallic hydrogen is expected to be realized in diamond anvil cells 
at pressures of order 400 GPa (currently pressures of around 320 GPa have already been 
reported on hydrogen \cite{2002}). The key issue centers on experimental
observability and  determining whether or not it is a liquid two-component
{\it superconducting superfluid}.  While standard superconductivity-detection procedures may be
inapplicable, we have proposed several alternative experimental probes and also 
pointed out their limitations. Keeping in mind that precise values of the physical parameters of the 
projected superfluid state are as yet unknown, we have based our analysis exclusively 
on topological properties in order to single out effects which are {\it qualitatively} 
different from the case where the system would be a  one-component electronic 
superconductor. The  possible experimental probes proposed here show that protonic 
superconductivity in a high pressure anvil cell {\it is experimentally accessible 
in principle} and therefore these probes should permit an answer to the question of 
the possible existence of two projected novel states of matter: the metallic and the 
superconducting superfluids.

This work was supported by STINT and the Swedish Research Council, the National Science 
Foundation, Grant No. DMR-0302347, the Research Council of Norway, Grant Nos. 158518/431, 
158547/431 (NANOMAT), 157798/432. We thank R. Hemley, W. Nellis, R.J. Rivers and W. Zurek 
for many useful discussions.

\end{document}